\documentclass[twocolumn,showpacs,aps,prl]{revtex4}
\usepackage{graphicx}
\usepackage{dcolumn}
\usepackage{amsmath}
\usepackage{latexsym}
\usepackage{morefloats}

\begin {document}


\title {Space-filling Percolation}
\author
{
Abhijit Chakraborty and S. S. Manna
}
\affiliation
{
\begin {tabular}{c}
Satyendra Nath Bose National Centre for Basic Sciences,
Block-JD, Sector-III, Salt Lake, Kolkata-700098, India \\
\end{tabular}
}
\begin{abstract}
      A region of two-dimensional space has been filled randomly with large number of growing circular discs allowing 
   only a `slight' overlapping among them just before their growth stop. More specifically, each disc grows from a nucleation 
   center that is selected 
   at a random location within the uncovered region. The growth rate $\delta$ is a continuously tunable parameter of the 
   problem which assumes a specific value while a particular pattern of discs is generated. When a growing disc overlaps 
   for the first time with at least another disc, it's growth is stopped and is said to be `frozen'. In this paper we 
   study the percolation properties of the set of frozen discs. Using numerical simulations we present evidence for the 
   following: (i) The Order Parameter appears to jump discontinuously at a certain critical value of the area coverage; 
   (ii) the width of the window of the area coverage needed to observe a macroscopic jump in the Order Parameter tends 
   to vanish as $\delta \to 0$ and on the contrary (iii) the cluster size distribution has a power law decaying functional 
   form. While the first two results are the signatures of a discontinuous transition, the third result is indicative 
   of a continuous transition. Therefore we refer this transition as a discontinuous-like continuous transition similar 
   to what has been observed in the recently introduced Achlioptas process of Explosive Percolation. It is also observed 
   that in the limit of $\delta \to 0$, the critical area coverage at the transition point tends to unity, implying the 
   limiting pattern is space-filling. In this limit, the fractal dimension of the pore space at the percolation point 
   has been estimated to be $1.42(10)$ and the contact network of the disc assembly is found to be a scale-free network.
\end{abstract}
\pacs {
       64.60.ah 
       64.60.De 
       64.60.aq 
       89.75.Hc 
}
\maketitle

\section {1. Introduction}

      In the recently introduced concept of ``Explosive Percolation'' (EP) it has been suggested that the nature of transition
   may be discontinuous in some percolation models \cite {EP}. This means that the associated Order Parameter, estimated by the size of 
   the largest cluster, should undergo a discontinuous change at the point of transition. In the context of percolation theory
   such a discontinuous transition can happen only when the largest cluster merges with the maximal of the second largest 
   cluster, which also has a macroscopic size \cite {Lee,Manna}. Though the original model of Explosive Percolation had been
   studied on Random Graphs \cite {EP}, later different versions of EP have been studied on the square lattices
   \cite {Ziff,Ziff1}, on scale-free networks \cite {Cho,Radicchi} and also for real-world networks \cite {Pan}. Recently it 
   has been shown that, though a class of EP models exhibit very sharp changes in their Order Parameters for finite 
   size systems and appear to exhibit discontinuous transitions, they actually have continuous transitions in the asymptotic 
   limit of large system sizes \cite {Riordan}. Here we propose and study a variant of the Continuum Percolation (CP) model 
   \cite {Lubachevsky} to exhibit a similar discontinuous-like continuous transition.

      The original CP model can be stated in the following way. There one finds the minimal density of Lilies, floating at random
   positions on the water surface of a pond, such that an ant will be able to cross the pond walking on the overlapping Lilies 
   when the radii of the Lilies have a fixed value $R$ \cite {Grimmett}. This phenomenon can also be described 
   as how the global connectivity is achieved in a Mobile ad hoc network (MANET) where each node represents a mobile phone, located
   at a random position, with a range of transmission $R$ \cite {MANET}. Depending on $R$ there exists a critical density of Lilies 
   or phones where the long range correlation sets in. It is well-known that in both versions of the CP the transition is continuous 
   and they belong to the same universality class of ordinary lattice percolation \cite {Stauffer}.

      In comparison, here we study the percolation problem in an assembly of growing circular discs. These discs are 
   released one at a time at random positions and they grow at a uniform rate so that at any arbitrary intermediate stage 
   different discs have different radii. In general, a slight overlap among them is allowed when a disc grows to overlap with 
   another one for the first time. This mutual overlap ensures that the global connectivity is achieved at a certain density 
   of discs. In the long time the pattern of discs cover the entire space. We are interested in the study of percolation 
   properties of this space-filling pattern which, to our knowledge, has not been studied yet. 

\begin{figure*}[top]
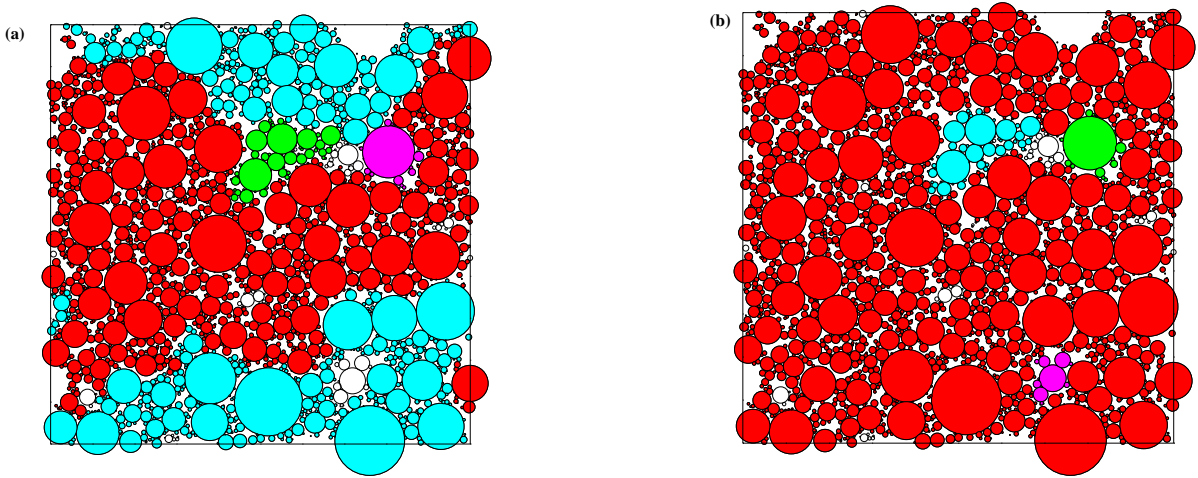

\begin {tabular}{cc}
\includegraphics[width=6.5cm]{figure01a.eps} \hspace*{1.5cm} & \hspace*{1.0cm}
\includegraphics[width=6.5cm]{figure01b.eps}
\end {tabular}
\caption{(Color online)
Top four largest clusters of the disc patterns (with $\delta = 0.001$) right before (a) and after (b) the maximal 
jump in the largest cluster. Different clusters have been shown by circles filled using different colors and rest 
of the circles are kept unfilled.
(a) At time $t = 1973$ the number of discs in the top four large clusters are 1167 (Red); 667 (Cyan); 57 (Green); 16 (Magenta),
(b) and at time $t = 1974$ they are 1836 (Red); 57 (Cyan); 16 (Green); 13 (Magenta).
}
\end{figure*}

      Various models of space-filling patterns have been studied in the literature characterized by their fractal dimensions
   $d_f$. In the Apollonian gasket, discs are placed iteratively in the curvilinear triangular spaces between the sets of 
   three mutually touching discs. Consequently the area of the uncovered space gradually vanishes and has the fractal dimension 
   $d_f=1.305686729(10)$ \cite {Thomas-Dhar}. In Space-filling bearing patterns a region of two dimensional space is covered 
   by an infinite set of mutually touching discs which can rotate without slipping with a fixed peripheral speed. Different 
   patterns have different fractal dimensions which vary between 1.3057 to 1.4321 \cite {Manna-Herrmann}. Due to their 
   deterministic algorithms the global connectivity of these patterns is guaranteed. On the other hand, in one model
   of random space-filling pattern of touching discs, such a global connectivity is not ensured. Here the nucleation centers of 
   the discs are selected at random locations in the uncovered region one after another. After introduction when 
   a disc grows all other discs remain frozen. Such a disc grows till it gets in contact with another 
   disc for the first time when it stops. Such a pattern has the fractal dimension $d_f \approx$ 1.64 \cite {MannaPhysica}.
   Recently space-filling patterns in three dimensional random bearings have been studied in \cite {Baram}.
   All these patterns, in the limit of infinite number of generations, are space-filling. It is known that, while this limit
   is being taken, the dust of remaining uncovered pore spaces form a Fractal set. The fractal dimension $d_f$ can be
   estimated following \cite {Herrmann}. 
   
      In section 2 we describe the model in detail, the data and their analysis are presented in section 3. Section 4 describes 
   the associated contact network and finally we conclude in section 5.

\section {2. The Model}

      The pattern of circular discs is generated within a unit square box placed on the $x-y$ plane. The radii of all 
   discs grow uniformly using a continuously tunable rate $\delta$. To generate a pattern, a specific value of $\delta$ 
   is assigned for all discs. The time $t$ is a discrete integer variable that counts the number of discs released.
   Therefore at time $t$ the pattern has exactly $t$ discs of many different radii. Initially the square box is completely 
   empty. Then at each time step a new disc with zero radius is introduced. Within the square box the 
   assembly of discs cover a region of space which is called the `covered region', the remaining space not covered by any of
   the discs is referred as the `uncovered region'. While growing, once a disc overlaps with another disc, it 
   stops immediately and does not grow any further. Such a disc is called a `frozen' disc.

      At any arbitrary intermediate time step the following activities take place: (i) A point is randomly selected 
   anywhere within the uncovered region. A circular disc of radius $\delta$ is placed with its center fixed at this point. 
   (ii) Simultaneously, the radii of all other non-frozen discs are also increased synchronously by the same amount $\delta$. 
   Every growing disc is checked if it has overlapped with any other disc, if so, it's growth is stopped and is declared as 
   a frozen disc.

      Gradually, the space within the square box is increasingly covered by the discs and therefore the amount of 
   uncovered area decreases with time. We define the control variable $p$ as the sum of the areas of all discs. It may be 
   noted that $p$ is slightly larger than the actual ``area coverage'' since the overlapped areas are doubly counted in the total 
   sum of disc areas. However it has been observed that the total overlap area tends to vanish in the limit of $\delta \to 0$ 
   and we would refer $p$ as the area coverage in the following discussion.

      In a particular run, the simulation is stopped only when the area coverage $p$ reaches a pre-assigned value or some pre-defined
   condition becomes valid. For example, to reach the percolation point, the run is terminated only when a global connectivity 
   appears through the overlapping discs from the top to the bottom for the first time. If one continues further, a stage would 
   come when the different pieces of uncovered regions would be so tiny that any newly introduced disc would freeze immediately 
   at the first time step. This would be the natural exit point of the simulation. However in most cases of our calculation
   simulations were run up to the percolation point.

      We are interested to study the percolation process of this growth model. Multiple overlapping discs form different
   clusters. Discs of a specific cluster are connected among themselves through overlaps. Size $s$ of a cluster is determined by
   the sum of the areas of all discs of the cluster. It has been observed, that in an arbitrary pattern near the percolation point, 
   typically there are two large clusters. For example in Fig. 1(a) we exhibit a typical disc pattern just before the percolation 
   point at time $t = 1973$ grown at a rate $\delta = 0.001$. Four top largest clusters are shown by discs, filled using different 
   colors. Right at the next time step $t$ = 1974 two small discs connect the top two largest clusters so that the size of the largest 
   cluster jumps from 0.439 to 0.747 (Fig. 1(b)). This behavior is typical of the percolation process studied here. Prior to the
   percolation point the largest and second largest clusters have a tendency to compete and grow simultaneously while maintaining 
   their comparable sizes. We define the percolation point when the Order Parameter jumps by a maximum amount in a single time
   step. This happens only when the largest cluster merges with the maximal of the second largest cluster. In the following we 
   present simulation results exhibiting this behavior.
   
      Given a specific growth rate $\delta$ one generates the disc assembly till the percolation point. At this stage a 
   probability distribution $n(r,\delta)$ of radii $r$ of the discs in the pattern is defined. Consequently, for the subset of discs 
   whose radii are at least $r$, one further estimates the cumulative distribution $N(r,\delta)$; the total perimeter $P(r,\delta)$ 
   of all discs in the subset and $A(r,\delta)$ as the total remaining uncovered area external to all discs in the subset. It is 
   assumed that in the limit of $r \to 0$ all these quantities vary as some powers of $r$ as follows \cite {Herrmann}:
\begin {align}
 n(r,\delta) & = \Sigma_{r_i = r}1 \sim r^{-(d_f+1)} \nonumber \\
 N(r,\delta) & = \Sigma_{r_i \ge r}1 \sim r^{-d_f} \nonumber \\
 P(r,\delta) & = 2\pi \Sigma_{r_i \ge r}r_i \sim r^{1-d_f} \nonumber \\
 A(r,\delta) & = 1 - \pi \Sigma_{r_i \ge r}r^2_i \sim r^{2-d_f} 
\end {align}

\begin{figure}
\begin {center}
\includegraphics[width=7.5cm]{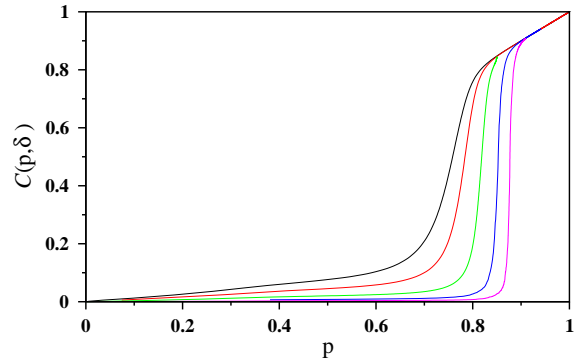}
\end {center}
\caption{(Color online) 
Order Parameter ${\cal C}(p,\delta)$ has been plotted against the area coverage $p$ for growth rates $\delta =$ 0.002 (black), 
0.0008 (red), 0.0002 (green), 0.00004 (blue) and 0.00001 (magenta).
}
\end{figure}

\section {3. The results}

      Let $s_m(p,\delta)$ denote the size (i.e., maximal covered area) of the largest cluster. Then the Order Parameter 
   ${\cal C}(p,\delta)$ of the growth process is determined by the average size of the largest cluster of the pattern for 
   an area coverage $p$, 
\begin {equation}
{\cal C}(p,\delta) = \langle s_m(p,\delta) \rangle
\end {equation}
   the average being taken over a large number of uncorrelated growth processes. Since no checking of the global connectivity 
   is required in this part of the simulation, we have used the periodic boundary condition along both the $x$ and $y$ axes. 
   In Fig. 2 we plotted ${\cal C}(p,\delta)$ with $p$ for five different values of the growth rate $\delta$. It has been observed 
   that for every plot that around a specific value of $p$ = $p_c(\delta)$ the growth of the Order Parameter is very sharp. This 
   happens because for a typical run $\alpha$ the maximal jump in $s^{\alpha}_m(p,\delta)$ takes place at $p^{\alpha}_c$. 

\begin{figure}[t]
\begin {center}
\includegraphics[width=7.5cm]{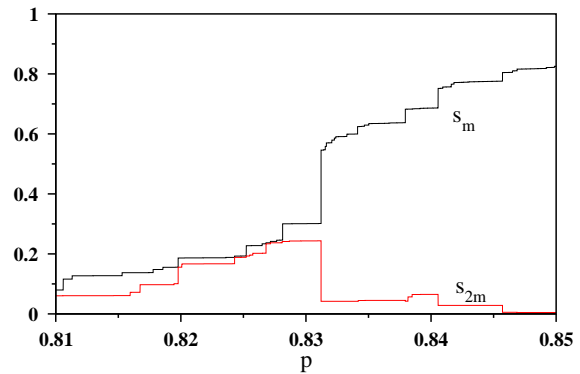}
\end {center}
\caption{(Color online) 
Variation of the sizes of the largest (black) cluster $s^{\alpha}_m$ and the second largest (red) cluster $s^{\alpha}_{2m}$ with the 
area coverage $p$ for a single run and for $\delta = 0.0001$. While $s^{\alpha}_m$ increases monotonically, $s^{\alpha}_{2m}$ increases 
to a maximum and then drops to a small value when the second largest cluster merges with the largest cluster. This corresponds to the maximal jump in the largest cluster and is identified as the percolation point for the $\alpha$-th run.
}
\end{figure}

      A closer look into the growth process reveals that this maximal jump in the largest cluster occurs only when the maximum of the 
   second largest cluster merges with the largest cluster. This has been exhibited explicitly in Fig. 3 where we plot the
   sizes of the largest cluster $s^{\alpha}_m$ and that of the second largest cluster $s^{\alpha}_{2m}$ with the area coverage $p$ 
   for a single run $\alpha$. While $s^{\alpha}_m$ grows monotonically, growth of $s^{\alpha}_{2m}$ is not so because it reaches
   to a maximum and then falls to a much lower value. We assume a nomenclature 
   that always the smaller cluster merges with the larger cluster. Therefore, when $s^{\alpha}_{2m}$ merges with $s^{\alpha}_m$ it is 
   the third largest cluster $s^{\alpha}_{3m}$ that becomes the second largest cluster $s^{\alpha}_{2m}$. This may happen a few times 
   and we mark that particular value of $p = p^{\alpha}_c$ where the maximal jump in $s^{\alpha}_m$ takes place.

\begin{figure}[t]
\begin {center}
\includegraphics[width=7.5cm]{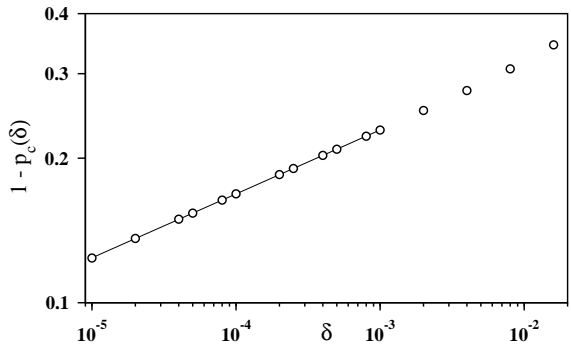}
\end {center}
\caption{
The deviation $1 - p_c(\delta)$ of the percolation threshold from unity has been plotted with the growth rate $\delta$ on a
$\log - \log$ scale. It is observed that as $\delta \to 0$ the deviation vanishes as a power of $\delta$ as:
$1 - p_c(\delta) \sim \delta^{0.133}$.
}
\end{figure}

      We define $p^{\alpha}_c(\delta)$ as the percolation threshold of the $\alpha$-th run \cite {Manna}. This value is 
   then averaged over a large number of un-correlated runs and the percolation threshold $p_c(\delta)$ is defined as:
\begin {equation}
p_c(\delta) = \langle p^{\alpha}_c(\delta) \rangle.
\end {equation} 
   It may be observed in Fig. 2 that the values of percolation thresholds $p_c(\delta)$ are gradually shifting towards 
   unity as $\delta \to 0$. To see this approach quantitatively, we plotted $1 - p_c(\delta)$ against $\delta$ on a 
   $\log - \log$ scale in Fig. 4. For small values of $\delta$ the data points indeed fit nicely to a straight line with 
   slope 0.133. Therefore we may write that as $\delta \to 0$,
\begin {equation}
1 - p_c(\delta) \sim \delta^{0.133}.
\end {equation}
   This implies that in the limit of $\delta \to 0$ the area coverage at the percolation point becomes unity. In other 
   words when the growth rate is infinitely slow, the global connectivity would appear for the first time when the entire 
   space would be covered by the discs and this limiting pattern is therefore space-filling. We therefore call this problem 
   as ``Space-filling percolation''.
   
      The percolation threshold has also been determined using the usual definition, i.e., when the global connectivity 
   appears for the first time in the system. In this case, periodic boundary condition along the $x$-axis 
   and open boundary condition along the $y$-axis have been used. For a specific run, as more and more discs are released, 
   we keep track if the connectivity between the top and the bottom boundaries of the unit square box through the system of 
   overlapping discs has appeared. When such a connectivity appears for the first time, we refer the corresponding pattern of 
   discs as the percolation configuration and define the total area coverage as the second definition of the percolation 
   threshold for this particular run. Like before, an average of these threshold values for a large number of independent 
   runs gives us the value of $p_c(\delta)$. It has been observed that the percolation thresholds measured using two methods 
   differ by small amounts, e.g., 0.039 and 0.019 for $\delta$ = 0.001 and 0.0001 respectively and this difference gradually 
   diminishes as $\delta \to 0$.

      At the percolation threshold $p^{\alpha}_c(\delta)$ of the $\alpha$-th run the disc pattern has one or more discs 
   which have  the largest radius $r^{\alpha}_m(\delta)$. The radius $\langle r_m(\delta) \rangle$ of the largest disc, 
   averaged over many runs, decreases as the growth rate $\delta$ decreases. A power law form along with a logarithmic 
   correction has been observed for this variation: 
\begin {equation}
\langle r_m(\delta) \rangle \sim \{\delta \log(1/\delta)\}^{0.331}.
\end {equation}
   In a similar way the average radius of a disc at the percolation point also has a power law variation as in the following.
\begin {equation}
\langle r(\delta) \rangle \sim \delta^{0.666}.
\end {equation}
   It may be noted that though $\langle r(\delta) \rangle < \langle r_m(\delta) \rangle$ the exponent of the former is larger 
   since as $\delta \to 0$ the value of $\langle r(\delta) \rangle$ decreases much faster than $\langle r_m(\delta) \rangle$.

      Eqn. (4) may be compared with the well-known relation of percolation theory that connects the ordinary site or bond percolation thresholds 
   $p_c(L)$ for finite size systems of length $L$ with the correlation length exponent $\nu$ as: $p_c(\infty) - 
   p_c(L) \sim L^{-1/\nu}$. In this case we may consider that the characteristic size of the objects filling the space is given by the 
   average $\langle r_m(\delta) \rangle$ of the maximal radius of the discs. Therefore for a given growth rate $\delta$ the system 
   size (equivalent to $L$), may be measured using $\langle r_m(\delta) \rangle$ as the yardstick and therefore 
   $L \sim 1/[\langle r_m(\delta) \rangle]$. Using Eqn. (4) and Eqn. (5) we can write:
 \begin {equation}
1 - p_c(\delta) \sim \delta^{0.133} \sim [\langle r_m(\delta) \rangle]^{0.133/0.331} \sim L^{-0.402}.
\end {equation}
   Using this equation we estimate $\nu = 0.331/0.133 \approx 2.49$ which is compared with $\nu = 4/3$ for ordinary
   percolation. We conjecture that in our case $\nu$ may be exactly equal to 5/2.

\begin{figure}[t]
\begin {center}
\includegraphics[width=7.5cm]{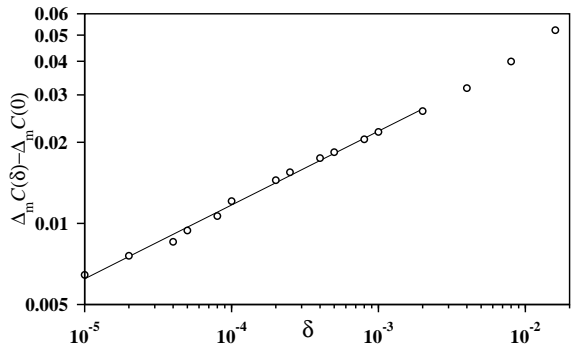}
\end {center}
\caption{
How the average value of the maximal jump in the Order Parameter 
$\Delta_m {\cal C}(\delta)$ approaches to $\Delta_m {\cal C}(0)$
has been shown. After some initial curvature the plot fits to a straight line as $\delta \to 0$.
This implies that Eqn. (8) indeed holds good with $\Delta_m {\cal C}(0)$ = 0.16 and $\mu = 0.274$.
}
\end{figure}

      Let us denote that the maximal jump in the Order Parameter by $\Delta_m {\cal C}(\delta)$. This is the average of the 
   maximal jumps in the size of the largest clusters over a large number of independent runs, i.e., 
   $\Delta_m {\cal C}(\delta) = \langle \Delta_{m} s_m(p,\delta) \rangle$. In Fig. 5 we plot $\Delta_m {\cal C}(\delta) - 
   \Delta_m {\cal C}(0)$ with $\delta$ on a $\log - \log$ scale. Since $\Delta_m {\cal C}(0)$ cannot be estimated directly
   we tried with different trial values of $\Delta_m {\cal C}(0)$ to make the plot which fits best to a straight line.
   Though there is an initial curvature for large values of $\delta$, the latter points obtained as $\delta \to 0$ fit 
   nicely to a straight line. This implies that the variation can be termed as a power law like:
\begin {equation}
\Delta_m {\cal C}(\delta) = \Delta_m {\cal C}(0) + A \delta ^{\mu}
\end {equation}
   with $\Delta_m {\cal C}(0) = 0.16$, $A = 0.15$ and $\mu = 0.274$. This relation can be interpreted that even in the limit of 
   $\delta \to 0$ the average maximal jump in the Order Parameter i.e., the area coverage of the largest cluster, is a finite 
   fraction of the entire area of the disc pattern. Therefore this is also another signature of the discontinuous percolation 
   transition in our model. 

      The rapidity with which the Order Parameter increases in Fig. 2 at the percolation threshold can also be quantified 
   by measuring the width of the window around the percolation threshold following the method used in \cite {EP}. For a
   single run, we define $p_{1/10}$ as the minimum value of the area coverage $p$ for which ${\cal C} > 1/10$. Similarly
   $p_{1/2}$ is the minimum value of $p$ for which ${\cal C} > 1/2$. The difference in these two area coverages $p_{1/2} 
   - p_{1/10}$ is the size of the window through which a 40 percent jump in the Order Parameter takes place. 

\begin{figure}[t]
\begin {center}
\includegraphics[width=7.5cm]{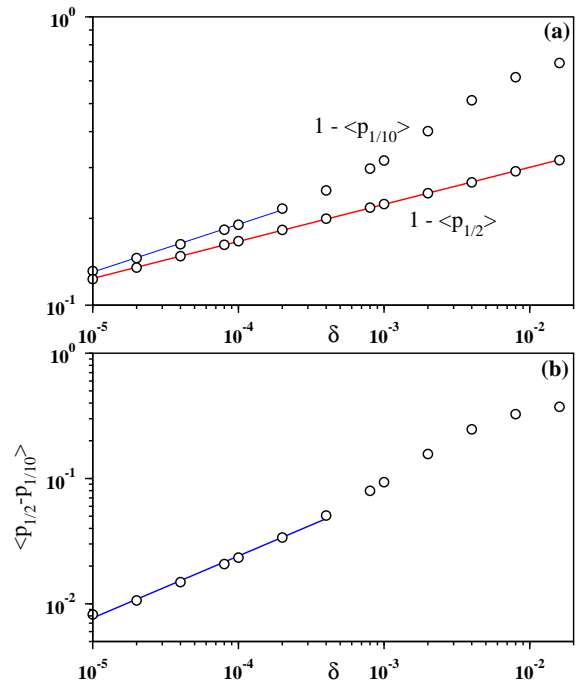}
\end {center}
\caption{(Color online) 
(a) Both $1 - \langle p_{1/10} \rangle$ and $1 - \langle p_{1/2} \rangle$ tends to vanish as $\delta^{0.166}$ and $\delta^{0.128}$ 
    respectively when $\delta \to 0$.
(b) Across the window $\Delta p = p_{1/2} - p_{1/10}$ the Order Parameter ${\cal C}(p,\delta)$ jumps from 1/10 to 1/2. 
    The average size of this window has been plotted with $\delta$ on $\log - \log$ scale and is observed to vanish as
    $\delta^{0.50}$ as $\delta \to 0$. 
}
\end{figure}

      Averaging over a large number of uncorrelated runs we estimated $\langle p_{1/10} \rangle$, $\langle p_{1/2} \rangle$
   and $\langle p_{1/2} - p_{1/10} \rangle$. In Fig. 6(a) we plotted $1 - \langle p_{1/10} \rangle$ and $1 - \langle p_{1/2} \rangle$.
   using a $\log - \log$ scale. It is observed that $1 - \langle p_{1/10} \rangle$ has some initial curvature for large values 
   of $\delta$ but as $\delta \to 0$ the curve become straight as a power law: $1 - \langle p_{1/10} \rangle \sim \delta^{0.166}$. 
   On the other hand $1 - \langle p_{1/2} \rangle$ fits to a nice straight line over the entire range of $\delta$ implying a
   power law variation $1 - \langle p_{1/2} \rangle \sim \delta^{0.128}$. In Fig. 6(b) we plotted $\langle p_{1/2} - p_{1/10} \rangle$
   against $\delta$. Here also, apart from some initial curvature for large $\delta$ the curve fits to a power law
\begin {equation}
\langle p_{1/2} - p_{1/10} \rangle \sim \delta^{0.50}.
\end {equation}
   Therefore as $\delta \to 0$ a 40 percent increase in the Order Parameter requires a vanishingly small change in the area coverage.
   This is again another evidence that the percolation transition in this model is likely to be a discontinuous transition.

\begin{figure}[t]
\begin {center}
\includegraphics[width=7.5cm]{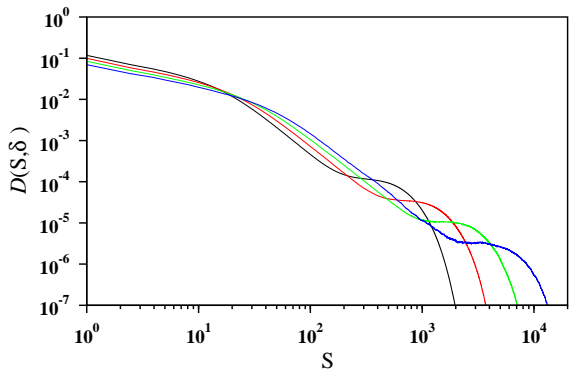}
\end {center}
\caption{(Color online) 
   Display of the binned data for the normalized cluster size distribution $D(S)$ measured by the number of discs $S$ in a cluster. 
   From the slopes in the intermediate regime we obtained $\tau$ = 2.16, 2.13, 2.14 and 2.13 respectively for growth rates $\delta$ 
   = 0.002 (black), 0.001 (red), 0.0005 (green) and 0.00025 (blue), $\delta$ decreases from left to right. The intermediate power 
   law regime gets elongated as $\delta$ decreases and we conclude a value of the associated exponent $\tau$ = 2.14(2).}
\end{figure}

      The percolation transition in our model is analyzed by yet another method, this time studying the cluster size 
   distribution at the percolation point. For this study we defined the cluster size $S$ in a different way, this time 
   it is the number of discs belonging to a specific cluster. In Fig. 7 we have presented the binned data for the 
   probability distributions of the cluster sizes for four different values of $\delta$. The cluster size distribution 
   data have been collected only when the global connectivity appears for the first time. All four $D(S,\delta)$ vs. 
   $S$ plots on the $\log$ - $\log$ scale have similar nature. After some initial slow variation, the $\log D(S,\delta)$ 
   decreases linearly with $\log S$ in the intermediate power law regime. Finally at the tail of the distribution there 
   is a hump, meaning an enhanced probability for the large clusters which connects the two ends of the system. It is 
   assumed and which seems to be very likely that as $\delta$ decreases the position of the hump shifts systematically 
   to large values of $S$ and therefore in the limit of $\delta \to 0$ the entire intermediate regime would fit to a power 
   law of the form: $D(S) \sim S^{-\tau}$. We conclude an average value of $\tau$ = $2.14(2)$ which is to be compared 
   with $\tau=187/91$ for ordinary percolation \cite {Stauffer}. This power law distribution of the cluster sizes can be 
   interpreted as an indication of a continuous transition for the percolation transition, in contradiction to the 
   discontinuous transition concluded from Eqns. (8) and (9).

      Here we recall that the original model of Explosive Percolation, which goes by the name of `Achlioptas Process' \cite {EP} has
   a similar story. For this model, most of the numerical results indicated that the associated percolation transition is 
   discontinuous. However, recently Riordan and Warnke have rigorously proved that a class of models using Achlioptas type
   processes are in fact continuous in the asymptotic limit of large size graphs \cite {Riordan}. We conclude that the 
   percolation transition in our model also behaves similarly to the Achlioptas Process, so that though apparently it 
   exhibits the behavior alike to a discontinuous transition, it is indeed a continuous transition. 

\begin{figure}[]
\begin {center}
\includegraphics[width=7.5cm]{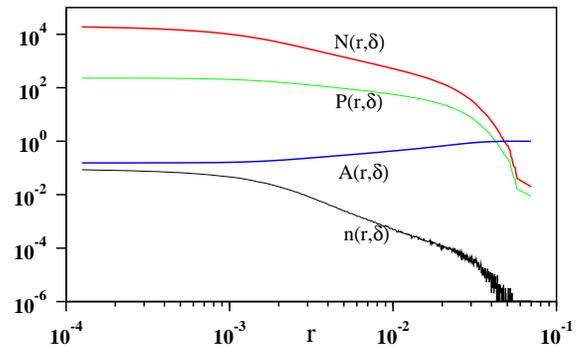}
\end {center}
\caption{(Color online)
The probability distribution $n(r,\delta)$ of the number of discs of radii $r$, its cumulative distribution $N(r,\delta)$, 
total perimeter $P(r,\delta)$ and the remaining uncovered area $A(r,\delta)$ for disc patterns grown at a rate $\delta = 
0.000125$. The slopes of each curves gives an estimate of the fractal dimension $d_f$ and an average value of 
$d_f = 1.42(10)$ has been obtained. 
}
\end{figure}

      Finally we measured the fractal dimension of the dust of pore spaces right at the percolation threshold using 
   the Eqn. (1). We considered a large sample of uncorrelated disc patterns that have been grown at a rate $\delta$. 
   A periodic boundary condition has been imposed along the $x$ direction and the pattern is grown till a global 
   connection appears along the $y$ axis when further growth is terminated. For each pattern we estimated the following 
   quantities: the distribution $n(r,\delta)$ of the number of discs of radii $r$, its cumulative distribution 
   $N(r,\delta)$, total perimeter $P(r,\delta)$ and the remaining uncovered area $A(r,\delta)$. We plot all four 
   quantities in Fig. 8 using a $\log - \log$ scale for $\delta = 0.000125$. For each curve the scaling appeared in 
   the intermediate regime of disc radii. Estimation of slopes of these curves in their scaling regions and using 
   Eqn. (1) we have obtained the values of the fractal dimensions as 1.42, 1.41, 1.40 and 1.46 respectively. Clearly 
   a large scatter of the estimated value of $d_f$ is present, yet this data indicates that the $d_f$ is like to be 
   around 1.42 a with rather large error of around 0.10. A more accurate estimation needs patterns to be generated 
   using even smaller value of the growth rate $\delta$.

\section {4. Contact Network}

      A contact network for the assembly of overlapping discs may be defined identifying the centers of the discs as 
   nodes. In addition a link between a pair of nodes is introduced if and only if their corresponding discs overlap 
   \cite {Herrmann1}. As time passes the contact network grows in the number of nodes as well as links. In Fig. 1(a) 
   we have exhibited the disc pattern at the percolation threshold where different discs are of different radii. In 
   general the large discs have overlaps with many other discs and therefore in the contact network these nodes form 
   the hubs of the network. In the same way smaller discs have fewer links but their numbers are more. The contact 
   network corresponding to the Fig. 1(a) has been exhibited in Fig. 9. Since there are many clusters, the network 
   is not a singly connected graph. The four top large clusters are represented by four sub-graphs of the network.

\begin{figure}[t]
\begin {center}
\includegraphics[width=6.5cm]{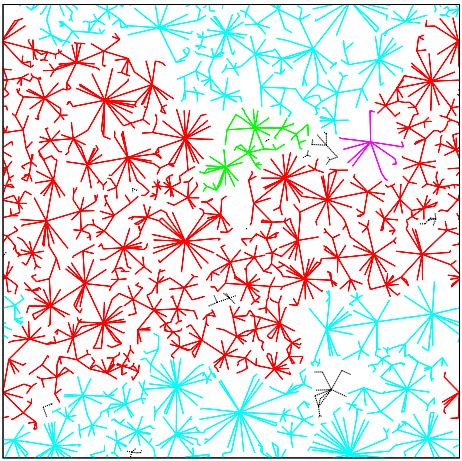}
\end {center}
\caption{(Color online) 
The picture of the contact network corresponding to the disc pattern in Fig. 1(a). A link between a pair of discs that
overlapped is drawn by a straight line joining their centers. The pattern has many clusters. The sub-graphs corresponding 
to the top four largest clusters are displayed by the same colors as used in Fig. 1(a).
}
\end{figure}

\begin{figure}[t]
\begin {center}
\includegraphics[width=7.5cm]{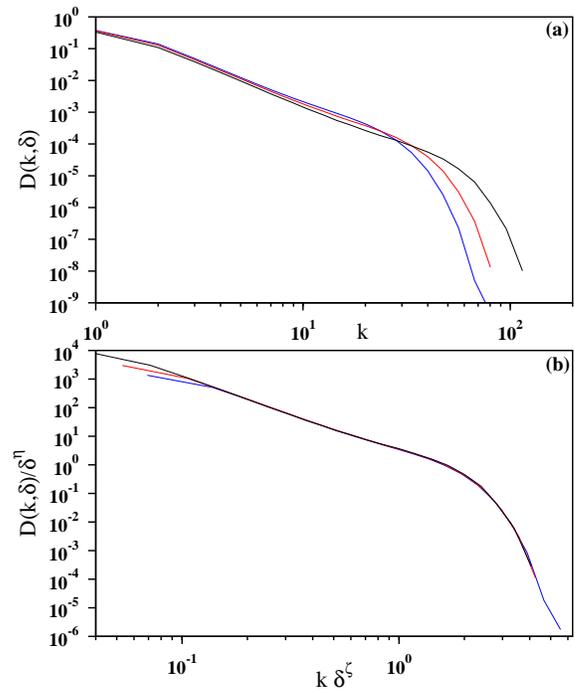}
\end {center}
\caption{(Color online) 
(a) The nodal degree distribution $D(k,\delta)$ of the contact network of the pattern at the percolation threshold
for three different values of the growth rate $\delta$ = 0.0001 (blue), 0.00004 (red) and 0.00001 (black). Binned 
data have been used to reduce fluctuation.
(b) The finite size scaling analysis of the data in (a) where $D(k,\delta)/\delta^{0.89}$ scales as $k\delta^{0.29}$
which gives $\gamma = 3.07$.
}
\end{figure}

      The degree $k_j$ of a node is the number of links meeting at node $j$. Here that would be equal to the number of other 
   $k_j$ discs that have overlap with the $j$-th disc. Typically such networks are scale-free networks which have power law 
   degree distributions. Similarly we expect that for the contact network of the disc pattern, the largest component of the 
   graph, right at the percolation threshold has the degree distribution $D(k) \sim k^{-\gamma}$, in the limit of $\delta 
   \to 0$ where $\gamma$ is the degree distribution exponent to be estimated. For finite $\delta$ we have calculated the 
   probability distribution $D(k,\delta)$ which is the probability that a randomly selected node has degree $k$. In Fig. 10(a) 
   we plot $D(k,\delta)$ against $k$ on a $\log - \log$ scale for the three different values of $\delta$. Apart from the very 
   small and large values of $k$ the curves are quite straight in the intermediate regimes indicating that the degree distributions 
   indeed have power law variations in the intermediate range of degree values. A direct measurement of slopes gives the values 
   of $\gamma(\delta)$ actually depend on $\delta$ and extrapolate like $\gamma(\delta) = \gamma(0) - 5.8\delta^{0.39}$ with 
   the extrapolated value of $\gamma(0)$ = 2.80. This analysis is supported by a finite size scaling analysis of the same data. 
   In Fig. 10(b) we plot the same data used in Fig. 10(a) but scale both the axes with suitable powers of $\delta$. The best data 
   collapse corresponds to 
\begin {equation}
D(k,\delta)/\delta^{\eta} \sim {\cal G}(k\delta^{\zeta})
\end {equation}
   where the values of $\eta \approx 0.89$ and $\zeta \approx 0.29$ are obtained; ${\cal G}(x)$ being the universal scaling function. 
   This implies that from the scaling analysis the estimate for the exponent $\gamma$ in the limit of $\delta \to 0$ is 
   $\gamma = \eta / \zeta \approx 3.07$. Averaging the two estimates, we quote a value of $\gamma = 2.94(14)$. 

      This estimated value of the degree distribution exponent close to 3, prompted us to compare this contact network with
   the Barab\'asi-Albert network \cite {Barabasi}. When a new disc is introduced, its center is selected randomly with uniform 
   probability anywhere within the uncovered region. For an already existing disc $i$ of radius $r_i$ if the center of the new 
   disc is selected within the annular ring of radius $(r_i+\delta)$ then the new disc becomes its neighbor immediately after 
   introduction. Therefore it is more probable that a newly added disc becomes the neighbor of a larger disc than a smaller one 
   and its probability is proportional to $r_i$. But this observation is until the disc $i$ is growing, i.e., till its degree 
   degree $k_i=0$. The moment it gets its first neighbor i.e., the first link, it's growth stops and it becomes frozen. Thereafter 
   the degree $k_i$ of $i$ increases but its radius does not, and gradually the annular space fills up. Therefore unlike the the 
   `rich gets richer' principle in the Barab\'asi-Albert network \cite {Barabasi} here the attachment probability is not 
   proportional to the degree $k$. We studied how the average radius of all discs whose degrees are equal to $k$ depends on $k$. 
   We have plotted (not shown here) for a small value of $\delta$ using the $\log$ - $\log$ scale $\langle r(k) \rangle /\log(k)$ 
   vs. $k$ on a $\log$ - $\log$ scale which fit very well to a straight line indicating that the following form with logarithmic 
   correction may be valid
\begin {equation}
\langle r(k) \rangle \sim k^{0.72}\log k
\end {equation}
   for the growth of the average radius of a disc with the their degree $k$.

\section {5. Conclusion}

      Signature of discontinuous jumps in the Order Parameter has been observed in a continuous percolation transition of 
   an assembly of growing circular discs which overlap slightly before becoming frozen. Extrapolation of numerical results
   indicate that in the limit of the extremely slow growth rate of $\delta \to 0$ the percolation transition occurs when
   the area coverage is unity, i.e., the disc pattern is space-filling even at the percolation point. Surprisingly, within 
   our numerical accuracy, it has been observed that the percolation transition of such a system has a discontinuous 
   macroscopic jump in the Order Parameter and also associated with a vanishing width of the transition window. On the 
   contrary, the cluster size distribution has been found to have a power law decaying functional form. We conclude that 
   though the transition in our model is actually continuous it exhibits certain features of a discontinuous transition. 
   We conclude that our model is yet another example like the Achlioptas Process \cite {EP}, the original model of Explosive 
   Percolation, where a similar discontinuous-like continuous transition has been observed.

\end {document}